\documentclass[aps,amssymb,twocolumn,superscriptaddress,nofootinbib]{revtex4}
\usepackage{setspace}
\usepackage{graphicx}
\usepackage{psfrag}
\def\RE {I\kern-6pt R    }
\def\Z  {Z\kern-13pt Z   }
\def\be {\begin{equation}}
\def\ee {\end{equation}  }
\def\beq{\begin{eqnarray}}
\def\eeq{\end{eqnarray}  }

\def\bi {\begin{itemize} }

\def\ei {\end{itemize}   }

\def\gtwid{\mathrel{\raise.3ex\hbox{$>$\kern-.75em\lower1ex\hbox{$\sim$}}}}
\def\ltwid{\mathrel{\raise.3ex\hbox{$<$\kern-.75em\lower1ex\hbox{$\sim$}}}}

\begin{document}


\title{Critical Behavior in the Gravitational Collapse of a Scalar Field
       with Angular Momentum in Spherical Symmetry}

\author{
	Ignacio (I\~naki) Olabarrieta
}
	\altaffiliation[Also at ]{TELECOM Unit, ROBOTIKER-Tecnalia, Ed. 202 Parque Tecnológico Zamudio E-48170, Bizkaia, Spain}
	\email{inakio@gmail.com}
\author{
	Jason F. Ventrella
}
	\email{ventrella@alum.mit.edu}
	\affiliation{
		Center for Computation and Technology, Louisiana State University, Baton Rouge, LA 70803 
	}
\author{
	Matthew W. Choptuik
}
	\altaffiliation[Also at ]
                  {Max-Planck-Institut f\"ur Gravitationsphysic, Albert-Einstein-Institut, Am M\"uhlenberg 1, D-14476 Golm, Germany}
	\email{choptuik@physics.ubc.ca}
\author{
	William G. Unruh
}
	\email{unruh@physics.ubc.ca}
	\affiliation{
		Department of Physics and Astronomy, University of British Columbia, Vancouver BC, V6T 1Z1 Canada \\
	}
	\affiliation{
      CIFAR Cosmology and Gravity Program
	}

\begin{abstract}
We study the critical collapse of a massless scalar field with angular momentum in spherical
symmetry.  In order to mimic the
effects of angular momentum we perform a sum of the stress-energy
tensors for all the scalar fields with the same eigenvalue $l$ of the angular momentum operator
and calculate the equations of motion for the radial part of these scalar fields.
We have found that the critical solutions for different values of $l$ are discretely
self-similar (as in the original $l=0$ case). The value of the discrete, self-similar period,
$\Delta_l$, decreases as $l$ increases in such a way that the critical solution 
appears to become periodic in the limit. The mass scaling exponent, $\gamma_l$, also decreases with $l$.
\end{abstract}



\maketitle

\section{Introduction}
\label{sec:introduction}

Most studies of black hole
critical phenomena 
(see \cite{Gundlach:1999cu}, \cite{Gundlach:2002sx} for reviews)
to date (or related phenomena in other sets of
nonlinear evolution equations)
have been
performed assuming spherical symmetry as a simplifying
assumption (exceptions are~\cite{Abrahams:wa},
\cite{Liebling:2002qp} and more recently
\cite{Choptuik:2003ac}).  This
simplification has been adopted in most cases because
accurate calculation of Type II critical solutions---which exhibit
structure at all scales due to their self-similar nature---requires great computational resources.
Since spherically symmetric spacetimes do not allow for angular
momentum, very little is currently known about the role
of angular momentum in critical collapse.  For a few
cases, most notably the Type II solutions found in spherically
symmetric collapse of a massless scalar field~\cite{Garfinkle:1998tt},
or certain types of perfect fluid~\cite{Gundlach:1997nb},~\cite{Gundlach:1999cw},
perturbative calculations about the spherical critical solutions
suggest that non-spherical modes, including those
contributing to net angular momentum, are damped as one
approaches criticality\footnote{There is some numerical evidence for growing {\em non-spherical} modes
in near-critical collapse, both for perfect fluids~\cite{Gundlach:1999cw} and massless scalar fields~\cite{Choptuik:2003ac}.
However, these modes appear to grow so slowly that, in both cases, it is expected that the spherical unstable mode continues to 
dominate near criticality.}. 
In particular in \cite{Garfinkle:1998tt},~\cite{Gundlach:1997nb}
and \cite{Gundlach:1999cw} using second order perturbation theory
it was predicted that the angular momentum
of the black holes produced should have the following dependence as a function of
the critical parameter $p$:
\begin{equation}
\label{L_scaling_law}
\vec L_{\rm{BH}} = \vec L_0 \left(p-p^\star \right)^\mu,
\end{equation}
where $\vec L_0$ is family-dependent and $\mu$ is a universal scaling exponent 
satisfying $\mu > 2\gamma$
($\gamma$ being the scaling exponent for the black hole mass).
Specifically, it was suggested that $\mu\approx0.76$ for the scalar field case, whereas the computations 
indicated that $\mu$ would depend on the equation of state for perfect fluid collapse.
These calculations thus suggest that, at least for small deviations
from spherical symmetry, the resulting solutions on the
verge of black hole formation should remain spherically symmetric in
non-symmetric collapse.
We also note that an axisymmetric numerical relativity
code has been developed~\cite{hlcp} to study non-perturbatively
some effects of angular momentum in the critical collapse of a scalar field.
Interestingly, the results found for $\Delta$ and $\gamma$ in the case of a complex scalar field 
with principal azimuthal ``quantum number'', $m=1$ 
are very close to the results we find in our model for $l=1$, as described in Sec.~\ref{sec:results}.

Here a different approach is taken. Maintaining spherical
symmetry, the equations of motion for a massless scalar field are
modified by effective terms which mock up some of the effects
of angular momentum. As described below, the procedure amounts to performing an angular average
over the matter field variables---similar to that done in \cite{Rein:1998uf},
\cite{Olabarrieta:2001wy} and \cite{Ventrella:2003fu}---and results in an entire family of models,
parameterized by a principal angular ``quantum number'', $l$ (we will
generally restrict $l$ to take on non-negative integer values, although
real-valued $l$'s are also formally possible).
We note that since the models remain spherically symmetric,
we cannot use them to address the validity of the perturbative calculations mentioned
above (e.g. equation (\ref{L_scaling_law})).
Nonetheless, we find interesting results that may shed some light on
the effects of angular momentum near the black hole threshold.
                                                                                                          
Some of the main results that have been found are as follows.
First, each value of the angular momentum parameter $l$ apparently defines a
distinct critical solution.
For $l < 10$, these solutions are found to be discretely self similar, with
values of the echoing exponent, $\Delta_l$, that rapidly decrease (approximately exponentially) as $l$
increases. As a result, for large values of $l$, and for the time scales
for which we are able to dynamically evolve near criticality, the threshold solutions
become approximately {\em periodic}.
In addition, and as expected for Type II solutions, we find that for
$l < 7$ the masses of the black holes formed follow power laws.
As with the echoing exponents, for increasing values of $l$ it is found that the
mass-scaling exponent, $\gamma_l$, rapidly decreases, again approximately exponentially in $l$.

The remainder of this paper is structured as follows. In the following
section we describe the recipe used to
calculate the effective equations of motion,
along with the regularity and
boundary conditions imposed in the solution of these equations.
In Sec.~\ref{sec:results} we briefly describe the
numerical code, the way the solutions have been analyzed, and then
provide a summary of the results obtained for varying values of $l$.
Throughout this paper we use units such that the universal gravitational constant, $G$, and
the speed of light in vacuum, $c$, are both unity.

\section{Equations of Motion}
\label{sec:equations}

\subsection{Equations}
In order to derive equations of motion,
scalar fields of the following form are considered:
\beq
\Psi^m_l(t,r,\theta,\phi) &=& \psi^{(l)}(t,r)\,Q_{lm}(\theta,\phi), \nonumber \\
  && m = -l, -l+1, \cdots, l-1, l,
\label{sep}
\eeq
where $Q_{lm}(\theta,\phi)$ are normalized
{\it real} eigenfunctions of the angular part of the flatspace
Laplacian with eigenvalue $l(l+1)$, and the index $m$ labels the
$2l+1$ distinct orthonormal eigenfunctions for a
given value of $l$.\footnote{Note that, in general,
$Q_{l m}(\theta, \phi)$ will {\em not} be eigenfunctions
of the azimuthal rotation operator $(\partial / \partial \phi)$
since they are real.} More explicitly:
\begin{equation}
Q_{l m} = \left\{ \begin{array}{l l}   Y_{l 0}  &{\rm for} \quad  m=0,\\
                \frac{1}{\sqrt{2}}  \left( Y_{l m}  + (-1)^m Y_{l -m}\right) & {\rm for} \quad m>0,\\
                \frac{1}{i\sqrt{2}} \left( Y_{l |m|}  -  (-1)^{|m|}Y_{l -|m|}\right) &{\rm for} \quad m<0,
                             \end{array}
                   \right.                   
\end{equation}
where $Y_{l m}\left(\theta, \phi \right)$ are the regular spherical harmonics.
By construction, the scalar fields $\Psi^m_l$ are not, in general,  spherically
symmetric and we therefore do not study
their collapse directly. Instead, our strategy is to
find effective equations for the {\em single} $(t,r)$-dependent
quantity $\psi^{(l)}(t,r)$, which we hereafter denote simply by $\psi$.  
To do so, for a specific value of $l$, we consider the
stress-energy tensors for the $2l+1$ fields $\Psi^m_l$:
\begin{equation}
\label{set_Tab}
T^{(l m)}{}_{ab} = \nabla_{a} \Psi^m_l \nabla_{b}\Psi^m_l
- \frac{1}{2}g_{ab}(\nabla^{c}\Psi^m_l\nabla_{c}\Psi^m_l),
\end{equation}
where $g_{ab}$ is the metric of the spacetime and $\nabla_a$ is
the metric-compatible covariant derivative.
Again by construction, and as is proven in Appendix~\ref{appendixA}, the sum of these stress tensors
\begin{equation}
{\mathcal T}^{(l)}{}_{ab}=\sum_m T^{(l m)}{}_{ab},
\end{equation}
{\em is} spherically symmetric, and thus depends only 
on $\psi(t,r)$, $l$,
and the metric $g_{ab}$. We can now compute the effective equation
of motion for the field, $\psi(t,r)$, using the fact 
that the divergence of the total stress energy tensor is zero, as is also proven in Appendix~\ref{appendixA}:
\begin{equation}
\label{conservation}
g^{ac}\nabla_{c}{\mathcal T}^{(l)}{}_{ab}= 0 \, .
\end{equation}
The equations for the geometric variables are determined from
the $3+1$ decomposition of the Einstein field equations.
For the current study we adopt
Schwarzschild-like (polar-areal) coordinates, in which the
metric takes the form:
\begin{eqnarray}
\label{polar-areal-metric}
ds^2=-\alpha^2(t,r)dt^2&+&a^2(t,r)dr^2\nonumber\\ &+& r^2d\theta^2 + r^2\sin^2{\theta}d\phi^2 \, .
\end{eqnarray}
Here $\alpha(t,r)$ is the lapse function and $a(t,r)$ is the
only non-trivial components of the 3-metric
(both $\alpha$ and $a$ are positive functions).
Using this metric, the non zero components of
the stress-energy tensor for a general value of $l$ are
\begin{eqnarray}
{\mathcal T}^{(l)}{}^t{}_t &=& -\frac{(2l+1)}{8 \pi}
            \left[\frac{1}{a^2}\left(\Pi^2 + \Phi^2 \right)
            +l(l+1)\frac{\psi^2}{r^2}\right],\\
{\mathcal T}^{(l)}{}^t{}_r &=& -\frac{(2l+1)}
           {8 \pi}\frac{2}{a \alpha}\,\Pi\,\Phi,\\
{\mathcal T}^{(l)}{}^r{}_r &=& \frac{(2l+1)}{8 \pi}
\left[\frac{1}{a^2}\left(\Pi^2 + \Phi^2 \right)
            -l(l+1)\frac{\psi^2}{r^2}\right],\\
{\mathcal T}^{(l)}{}^\theta{}_\theta &=&
{\mathcal T}^{(l)}{}^\phi{}_\phi =
\frac{(2l+1)}{8 \pi a^2}\left(\Pi^2 - \Phi^2 \right) \, ,
\end{eqnarray}
and the stress-energy trace is
\begin{eqnarray}
\label{trace_t}
{\mathcal T}^{(l)}&\equiv&
     {\mathcal T}^{(l)}{}^i{}_i \nonumber \\
     &=& \frac{(2l+1)}{8\pi}\left[\frac{2}{a^2}\left(\Pi^2-\Phi^2\right)- 
               2l(l+1)\frac{\psi^2}{r^2}\right]\! .
\end{eqnarray}
In the above expressions, we have made use of the auxiliary variables, $\Phi$ and
$\Pi$, defined as follows:
\begin{eqnarray}
\label{pp_cons}
\Phi\left(t,r\right)&=&\frac{\partial \psi}{\partial r}, \\
\Pi\left(t,r\right)&=&
\frac{a}{\alpha}\frac{\partial \psi}{\partial t}.
\end{eqnarray}
The dynamical equations of motion for these fields, which follow from the definition
of $\Phi$ as well as the wave equation for $\psi$ (which in turn can be derived
from the vanishing of the divergence of the total stress tensor (\ref{conservation})) are then:
\begin{eqnarray}
\label{pp_evo}
\frac{\partial\Phi}{\partial t}&=&
\frac{\partial}{\partial r}\left(\frac{\alpha}{a}\Pi\right),\\
\label{pi_evo}
\frac{\partial\Pi}{\partial t}&=&
\frac{1}{r^2}\frac{\partial}{\partial r}\left(r^2 \frac{\alpha}{a}\Phi\right)-
l(l+1)a\alpha \frac{\psi}{r^2}\, .
\end{eqnarray}

Note that the dependence of these equations on $l$ is only through
the last term in equation (\ref{pi_evo}) which is proportional to $l(l+1)/r^2$.
This term can be thought of as the field-theoretic extension of an analogous term
due to the angular momentum potential, $l^2/r^2$, in the 1-dimensional reduced
problem of a particle moving in a central potential.
                                                                                                          
As mentioned above, equations for the geometric variables result from
the $3+1$ decomposition of the field equations, as well as from our choice of coordinates.
Specifically, we have the following
\begin{eqnarray}
\label{hc_chp2}
\frac{1}{a}\frac{\partial a}{\partial r}&=&
\frac{(2l+1)}{2}r\left(\Pi^2+\Phi^2+l(l+1)\frac{a^2}{r^2}\psi^2\right)\nonumber\\
&-&\frac{a^2-1}{2 r},\\
\label{sc}
\frac{1}{\alpha}\frac{\partial \alpha}{\partial r}&=&
\frac{(2l+1)}{2}r\left(\Pi^2+\Phi^2-l(l+1)\frac{a^2}{r^2}\psi^2\right)\nonumber\\
&+&\frac{a^2-1}{2 r},\\
\label{a_evo}
\frac{\partial a}{\partial t}&=&(2l+1) r \alpha \Pi \Phi.
\end{eqnarray}
Equation (\ref{hc_chp2}) is the Hamiltonian constraint, which is used to determine the 3-metric
component, $a$. Similarly, the slicing condition~(\ref{sc}) fixes the
lapse function~$\alpha$ at each instant of time, and is often known as the
{\em polar slicing condition}. It can be derived from the demand that
${\rm Tr}\left(K_{ab}\right)=K^r{}_r + K^\theta{}_\theta + K^\phi{}_\phi =
K^r{}_r + 2K^\theta{}_\theta = 0$, for all times.
The Hamiltonian constraint and slicing condition, with appropriate regularity
and boundary conditions, completely fix the geometric variables in this coordinate
system. Equation (\ref{a_evo}) is an extra equation derived from the definition
of $K^r{}_r$ and the momentum constraint.
In our numerical solutions, it is used as a gauge of the accuracy of our calculations,
as well as to provide a replacement for the Hamiltonian constraint in certain strong
field instances where the numerical constraint solver fails.
In addition, we compute the mass aspect function, $M(t,r)$,
\begin{equation}
M(t,r) = \frac{r}{2}\left(1-\frac{1}{a^2} \right),
\end{equation}
which serves as a valuable diagnostic quantity in our calculations.
The value of this function as $r\to\infty$ agrees with the ADM mass, and more
generally, in a vacuum region of spacetime, measures the amount of (gravitating)
mass contained within the 2-sphere of radius $r$ at time $t$.
Moreover, $2M(t,r)/r$ is useful since its value
approaches $1$ when a trapped surface is developing and hence (modulo cosmic censorship),
a black hole would form in the spacetime being constructed.
We note that, as is the case with the usual Schwarzschild coordinates for a
spherically symmetric black hole, polar-areal coordinates cannot penetrate
apparent horizons, and in fact become singular as they come ``close to''
black-hole regions of spacetime, where $2M(t,r)/r \to 1$.
This fact does not present a problem in the study of critical behavior in
our models, since the critical solutions {\em per se} have $\max_r\left\{2M(t,r)/r\right\}$
bounded away from $1$.

\subsection{Regularity and  Boundary Conditions}
In addition to the above equations of motion, appropriate regularity and boundary
conditions are needed. At the origin,
$r=0$, regularity is enforced via
\begin{eqnarray}
a(t,0) &=&1,\\
\frac{\partial a}{\partial r}(t,0) &=&0,\\
\frac{\partial \alpha}{\partial r}(t,0) &=&0,\\
\label{psi_fall_off}
\psi(t,0) &=&O(r^l),\\
\label{pi_fall_off}
\Pi(t,0)  &=&O(r^l),\\
\label{pp_fall_off}
\Phi(t,0) &=& \left\{
              \begin{array}{lll}
                O(r^{l-1}) & \hbox{\rm for} & l \ge 1,\\
                O(r)       & \hbox{\rm for} & l=0.
              \end{array} \right.
\label{phir0}
\end{eqnarray}
In the continuum, our equations of motion are to be solved as a pure Cauchy problem,
on the domain $t \ge 0$, $r \ge 0$, with boundary conditions at spatial infinity
given by asymptotic flatness (i.e.~that the matter fields vanish, and that the metric
becomes that of Minkowski spacetime, as $r\to\infty$). Computationally, we solve
an approximation to this problem on a finite spatial domain $0 \le r \le r_{\rm max}$,
where $r_{\rm max}$ is some arbitrary outer radius chosen sufficiently large that
we are confident that the numerical results do not depend significantly on its
precise value.
At the outer boundary, then, the following
condition for $\alpha$ is imposed:
\begin{equation}
\label{alpha_bc}
\alpha(t,r_{\rm max})\ a(t,r_{\rm max})=1.
\end{equation}
This can be viewed as simply providing a convenient normalization for $\alpha$,
since given a solution, $\alpha$,  of the slicing equation~(\ref{sc}), $k \alpha$
is also a solution, where $k$ is an arbitrary positive constant. We note
that although we have used (\ref{alpha_bc}) in order to perform the calculations,
a different normalization convention---i.e.~a different, and time dependent,
choice of $k$---has been used in order to perform the analysis of the solutions.
Specifically, in the analysis we have used central proper time $T$ defined by:
\begin{equation}
\label{central_t}
T = \int^{T}_0 \alpha({\tilde t},0) \, d{\tilde t}\,.
\end{equation}
This definition of time has a natural geometrical interpretation since $r=0$ is invariantly
defined by the symmetry of the spacetime.
For the scalar field variables, $\Pi$ and $\Phi$,
approximate outgoing-radiation boundary conditions (Sommerfeld conditions) are used:
\begin{eqnarray}
\frac{\partial \Phi}{\partial t}(t,r_{\rm max})+
\frac{\partial \Phi}{\partial r}(t,r_{\rm max})
+\frac{\Phi(t,r_{\rm max})}{r_{\rm max}}&=&0,\\
\frac{\partial \Pi}{\partial t}(t,r_{\rm max})+
\frac{\partial \Pi}{\partial r}(t,r_{\rm max})
+\frac{\Pi(t,r_{\rm max})}{r_{\rm max}}&=&0.
\end{eqnarray}
                                                                                                          
An important point in the derivation of the equations of motion is the
fact that the eigenfunctions in~(\ref{sep}) are discrete and the allowable values of $l$ are
only non-negative integers. Once the equations are obtained we have relaxed
that constraint and have allowed $l$ to take non-negative {\em real} values. The solutions
corresponding to non-integer values of $l$ would have some degree
of irregularity at the origin depending on the particular value of $l$ chosen.
This implies that only some finite number of derivatives with respect to $r$
will be defined at $r=0$. In our particular numerical implementation, which assumes that
second derivatives of the variables are defined,
we have been able to study the evolution of these systems as long as $l>3$.

\section{Results}
\label{sec:results}

\subsection{Numerics}
\label{numerics}
We solve
equations (\ref{pp_evo}), (\ref{pi_evo}) for the scalar field gradients,
equations (\ref{hc_chp2}), (\ref{sc}) for the geometry,
and use (\ref{pp_cons}) to reconstruct the field $\psi$.
The system is approximated using
second order centered finite difference techniques, and coded
using RNPL~\cite{rnpl}.
Numerical dissipation of the Kreiss-Oliger~\cite{KO} variety
was included to damp high frequency modes, and it should be noted
that this particular type of dissipation is added at sub-truncation
error order, so does not affect the overall accuracy of the scheme
as the mesh spacing tends to 0.
For the current computations, the damping terms were most useful in
regularizing the truncation error estimation procedure that occurs
when adaptive mesh refinement (AMR) techniques are used.
It was also crucial to impose the correct
leading-order regularity conditions close to the origin, $r=0$ (equations
(\ref{pi_fall_off})-(\ref{pp_fall_off})), in order to keep the solution
regular during the evolutions. Most of the calculations were done
on a fixed uniform spatial grid $r_j = (j - 1)\Delta r$, $j = 1, 2, \cdots, J$,
$J = 1 + r_{\rm max} / \Delta r$ with a typical number of grid points $J=1025$,
and the outer boundary of the computational domain typically at $r_{\rm max}=100$.
For small values of the angular momentum parameter---specifically for
$l\le2$---an AMR algorithm based on that described in~\cite{Choptuik:jv}
was used.

\subsection{Families of Initial Data}
Our study involved the evolution of $6$ different one
parameter families of initial data, each defined by an initial profile
$\psi(0,r)$ as listed in Table
\ref{families_id_table}, with specific values of the parameters
appearing in the profile definitions
as given in Table \ref{initial_data_table}.
In addition to $\psi(0,r)$, we need to provide $\Pi(0,r)$ to complete the  specification of the initial data. 
In all cases we chose $\Pi(0,r)$ to produce an approximately in-going pulse at the initial time:
\begin{equation}
\Pi(0,r) = \Phi(0,r) = \frac{\partial \psi}{\partial r}(0,r).
\end{equation}

\begin{table}[]
\begin{center}
\begin{tabular}{ccc}
\hline
 Family  & Form of initial data, $\psi(0,r)$                                   & $p$  \\ \hline
 (a)     & $A\,\exp\left(-(r-r_0)^2/\sigma^2\right)$                        & $A$  \\ \hline
 (b)     & $-2 A\,(r-r_0)/\sigma^2\exp
\left(-(r-r_0)^2/\sigma^2\right)$     & $A$  \\ \hline
 (c)     & $A\,r^2\left( {\mathrm{atan}}(r-r_0)-
{\mathrm{atan}}(r-r_0-\sigma) \right)$ & $A$  \\ \hline
\end{tabular}
\end{center}
\caption[Families of initial Data.]{Families of initial data and the parameter $p$ that is tuned to generate a critical solution.}
\label{families_id_table}
\end{table}

As previously mentioned, all of the initial data families listed
in Table~\ref{families_id_table} have a single
free parameter, $p$, and, as is the usual case in studies of
black hole critical phenomena, for any given family we
observe two different final states in the evolution, depending
on the value of $p$.
For values of $p > p^\star$ the maximum value of $2M(t,r)/r$
approaches $1$ implying that an apparent horizon is about to form.
On the other hand if $p < p^\star$ the scalar field completely
disperses, and leaves (essentially) flat spacetime in its wake.
The solution that arises as $p \to p^\star$ then represents the
threshold of black hole formation and, by definition, is the
critical solution.  We note that these critical solutions are
{\em not} $t\to\infty$ end-states of evolution; rather they
persist for only a finite amount of time, and, in fact, are
unstable, heuristically representing an infinitely fine-tuned balance between
dispersal and gravitational collapse.

\begin{table}[]
\begin{center}
\begin{tabular}{ccc}
\hline
 Initial Data (F) & Family & Parameters                  \\ \hline
 1           & (a)    &$r_0=70.0$,\ \ $\sigma=5.00$  \\ \hline
 2           & (b)    &$r_0=70.0$,\ \ $\sigma=5.00$  \\ \hline
 3           & (c)    &$r_0=70.0$,\ \ $\sigma=5.00$  \\ \hline
 4           & (a)    &$r_0=40.0$,\ \ $\sigma=10.0$  \\ \hline
 5           & (a)    &$r_0=40.0$,\ \ $\sigma=5.00$  \\ \hline
 6           & (a)    &$r_0=70.0$,\ \ $\sigma=10.0$  \\ \hline
\end{tabular}
\end{center}
\caption[Type of initial data.]
{Initial data used in our investigations. The family labels are defined
in Table I. }
\label{initial_data_table}
\end{table}

\subsection{Analysis}
We have calculated $p^\star$ for the different families of
initial data described above, and for different values of $l$, via bisection (binary search),
tuning $p$ in each case to a typical precision of
$\left(p-p^\star\right)/p\approx 10^{-15}$ (which is close to
machine precision using 8-byte real floating point arithmetic).
                                                                                                
As in the case for $l=0$ (where the equations of motion reduce to those
for a single, non-interacting massless scalar field, as studied in~\cite{Choptuik:jv}),
the critical solutions for values of
$l\le 9.5$ are apparently discretely self similar (DSS). DSS spacetimes are scale-periodic, 
meaning that any non-dimensional quantity, $Z$, obeys the following
equation for some specific values of the parameters $\Delta$ and
$T^\star$:
\begin{equation}
\label{scaling}
Z\left((T-T^\star),r\right) =
Z\left(e^{n \Delta} (T-T^\star),e^{n \Delta}r\right),
\end{equation}
where $T$ is central proper time as defined by (\ref{central_t}), and $T^\star$ is
the ``accumulation time'' of the self-similar solution.
In (\ref{scaling}) the integer $n$ denotes the ``echo'' number.
We also note that due to the discrete $\psi \to -\psi$ invariance
that is exhibited both by the equations of motion as well as the critical
solutions themselves, if $\Delta$ is the echoing exponent for which formula
(\ref{scaling}) is satisfied with $Z(T,r) \equiv \psi(T,r)$,
then the geometric quantities $a(T,r)$, $\alpha(T,r)$, $2 M(T,r) / r$
obey (\ref{scaling}) with an echoing exponent $\Delta/2$.
                                                                                         
In order to extract $\Delta$ from our calculations, we use the
observation that certain geometric quantities will achieve
(locally) extremal values on the spatial domain at discrete central proper times
 $T_n$ given by
\begin{equation}
T_n-T^\star = \left(T_0 - T^\star \right) e^{n \Delta/2 }
\end{equation}
where $T_0$ is the time at which one starts counting the echoes.
Specifically, $\Delta$ and $T^\star$ have been computed by
a least squares fit for the times $T_n$ at
which $\max_r\left\{2 M(t,r)/r\right\}$ achieves a local maximum in time,
i.e.~by minimizing:
\begin{equation}
\chi^2=\sum_{n=1}^{N} \left\{T_n - T_0 e^{n\Delta /2}+
T^\star \left(e^{n \Delta /2}-1 \right)\right\}^2.
\end{equation}
\begin{table}[htb]
\begin{center}
\begin{tabular}{|c|c|c|}
\hline
 $l$ & $\Delta_l$           & $\gamma_l$          \\ \hline
0    & 3.43   $\pm$ 0.05    & 0.376  $\pm$ 0.003  \\ \hline
1    & 0.460  $\pm$ 0.002   & 0.119  $\pm$ 0.001  \\ \hline
2    & 0.119  $\pm$ 0.003   & 0.0453 $\pm$ 0.0002 \\ \hline
3    & 0.039  $\pm$ 0.001   & 0.020  $\pm$ 0.001  \\ \hline
3.5  & 0.0224 $\pm$ 0.0009  & 0.0127 $\pm$ 0.0008 \\ \hline
4    & 0.0132 $\pm$ 0.0008  & 0.0082 $\pm$ 0.0008 \\ \hline
4.5  & 0.0077 $\pm$ 0.0007  & 0.0052 $\pm$ 0.0006 \\ \hline
5    & 0.0044 $\pm$ 0.0007  & 0.0033 $\pm$ 0.0005 \\ \hline
5.5  & 0.0026 $\pm$ 0.0006  & 0.0020 $\pm$ 0.0005 \\ \hline
6    & 0.0015 $\pm$ 0.0005  & 0.0013 $\pm$ 0.0005 \\ \hline
6.5  & 0.0009 $\pm$ 0.0005  & 0.0008 $\pm$ 0.0005 \\ \hline
7    & 0.0006 $\pm$ 0.0004  &           -         \\ \hline
7.5  & 0.0004 $\pm$ 0.0004  &           -         \\ \hline
8    & 0.0003 $\pm$ 0.0004  &           -         \\ \hline
8.5  & 0.0002 $\pm$ 0.0003  &           -         \\ \hline
9    & 0.0002 $\pm$ 0.0004  &           -         \\ \hline
9.5  & 0.0002 $\pm$ 0.0003  &           -         \\ \hline
\end{tabular}
\end{center}
\caption[
Properties of critical solutions for different values of $l$.
]{Summary of the properties of the critical solutions computed for different
values of $l$. Note that both the echoing exponents, $\Delta_l$,
and the mass scaling exponents, $\gamma_l$, rapidly decrease as $l$ increases.
Quoted errors have been estimated from the variation in values computed
across the different families of initial data.
Values of $\Delta_l$ have been calculated using central proper time normalization
of the lapse function, which is the natural normalization for type-II critical
behavior. For
$l>6.5$ we have not been able to calculate $\gamma_l$ due to lack of numerical
precision.  Note that the $l=0$ data agree with
the original values calculated in \cite{Choptuik:jv}, and that the $l=1$ data agree
with values calculated in \cite{Liebling:1999ke} and \cite{Husa:2000kr}
using models of completely different origin.}
\label{delta_table}
\end{table}
\subsection{Results}
\begin{figure}
\begin{center}
\includegraphics[width=8.0cm,clip=true]{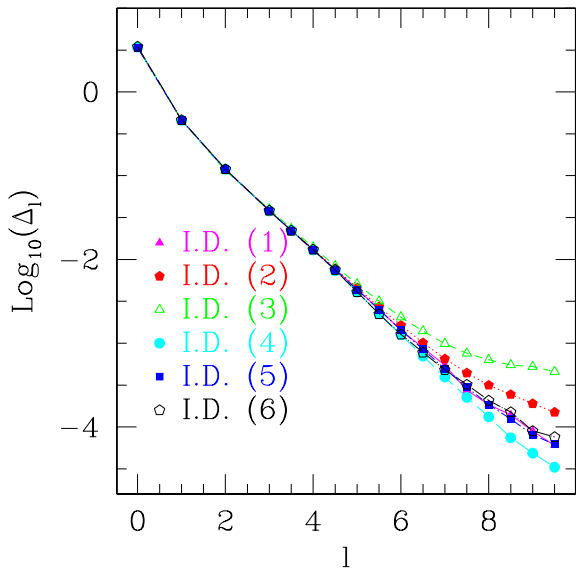}
\end{center}
\caption[Values of ${\mathrm{log}_{10}}\left(\Delta_l\right)$ versus $l$.]
{
Values of ${\mathrm{log}_{10}}\left( \Delta_l \right)$ versus $l$.
In this figure we can see that $\Delta_l$ decreases almost
exponentially with $l$. The different lines represent different families of
initial data. Assuming universality, the differences between the values
calculated for the different families provides
one measure of error in our determination of $\Delta_l$.
}
\label{deltas_fig}
\end{figure}

Table \ref{delta_table} summarizes the values of $\Delta_l$
we have estimated using this procedure; the data are also graphed in 
Fig.~\ref{deltas_fig}.  Again, note that the reported values for $\Delta_l$ 
have been calculated using central proper time $T$ instead of proper time at 
infinity (the parameterization used in the numerical evolutions {\em per se}).
Also the reported uncertainties have been estimated from the deviations in 
the $\Delta_l$ values computed across the the six different families of initial data.
The first entry in Table \ref{delta_table} ($l=0$) corresponds to the original case 
studied in \cite{Choptuik:jv}. The second one ($l=1$) is apparently the same solution 
found for the self-gravitating collapse of an $SO(3)$ non-linear $\sigma$ model, 
assuming a hedgehog 
ansatz~\cite{Husa:2000kr},~\cite{Liebling:1999ke}. 
Interestingly, the values for $\Delta_1$
and $\gamma_1$ also agree quite well with the values obtained from the study of the axisymmetric collapse 
of a complex-valued scalar field with azimuthal quantum number $m=1$~\cite{hlcp}, where values $\Delta \approx 0.42$ and 
$\gamma \approx 0.11$ are quoted.  However,
in the model considered in~\cite{hlcp}, the overall solution is clearly different because it is 
{\em not} spherically symmetric. The remainder of the solutions (for the other values of $l$) 
are, to the best of our knowledge, new.

\begin{figure}
\begin{center}
\includegraphics[width=8.0cm,clip=true]{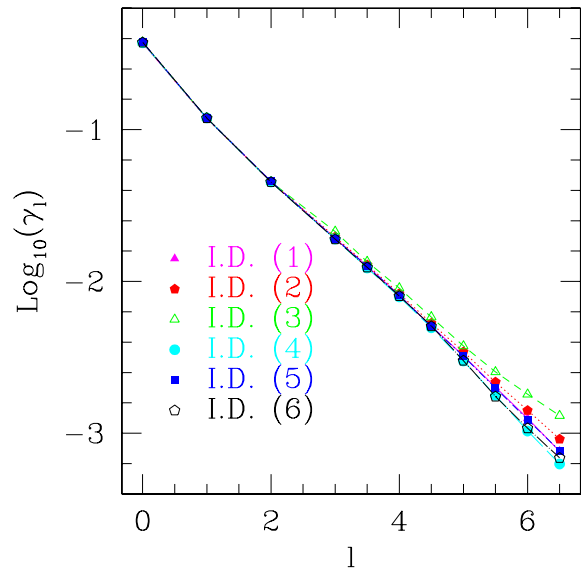}
\end{center}
\caption[Values of ${\mathrm{log}_{10}} \left( \gamma_l \right)$ versus $l$.]
{Values of ${\mathrm{log}_{10}}\left( \gamma_l \right)$ versus $l$, where
$\gamma_l$ is the scaling exponent defined by~(\ref{mass-scaling}).
As for the case of the echoing exponent, $\Delta_l$, $\gamma_l$ also decreases approximately
exponentially with $l$. We note that due to lack of numerical
precision we can only reliably compute $\gamma_l$ for $l\le6.5$
}
\label{gammas_fig}
\end{figure}

Systems exhibiting type II critical behavior, where the critical solution is 
self-similar, generally also exhibit power-law scaling of dimensionful quantities in 
near-critical evolutions.  For example, we can expect the black hole mass, $M_{\rm BH}$,
to scale as
\begin{equation}
\label{mass-scaling}
M_{\rm BH} \sim C \left(p-p^\star \right)^{\gamma_l}
\end{equation}
for super-critical evolutions as $p\to p^\star$~\footnote{In accord with the results in~\cite{hod} 
and~\cite{Gundlach:1997}, we expect small amplitude oscillations with period $\Delta$ to be superimposed
on the scaling law~(\ref{mass-scaling}).  We have, however, made no attempts to measure this effect in 
the current work.}.
Here $C$ is a constant that depends on the family of initial data while
$\gamma_l$ is a universal exponent for each value of $l$, i.e.~independent of the 
specific initial data family used to generate the critical solution.
We have observed such scaling in at least some of our computations, but,
following Garfinkle and Duncan~\cite{Garfinkle:1998va} have
found it more convenient to extract $\gamma_l$ by monitoring the maximum value
of the trace of the stress tensor, ${\cal T}$, which, from the Einstein equations, is
proportional to the maximum value of the Ricci curvature. On dimensional
grounds $\cal T$ (defined by~(\ref{trace_t})) and $R$ should both scale
with an exponent $-2 \gamma$.
This technique has the advantage of being more precise than a strategy
based directly on~(\ref{mass-scaling}) since
we can calculate the trace of the stress-energy
more accurately than the mass of the black hole formed, and can perform
the computation using sub-critical evolutions, where
the gradients of field variables generally do not become as large
as those in the super-critical cases. The values of $\gamma_l$ as
a function of $l$ are listed in Table~\ref{delta_table} and are
plotted in Fig.~\ref{gammas_fig}.

\begin{figure}
\begin{center}
\includegraphics[width=8.0cm,clip=true]{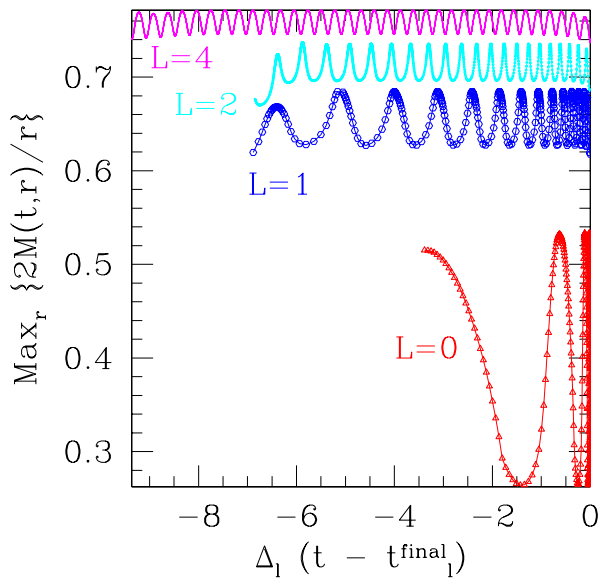}
\end{center}
\caption[Time evolution of the Maximum in $r$ of $2M(t,r)/r$ for $l=0$, $l=1$, $l=2$ and $l=4$.]
{
Evolution in time of the maximum in $r$ of the function $2M(t,r)/r$ for four
different critical solutions with increasing value of $l$ ($l=0$, $l=1$, $l=2$ and $l=4$).
The plot shows the evolution during the period of time when each solution shows 
discrete self similarity. The time coordinate is rescaled by $\Delta_l$ for visualization
purposes and is shifted so that the function values coincide at $t=0$. We note how 
the solutions tend to periodicity 
with increasing
values of $l$.
}
\label{raw_data}
\end{figure}

As is characteristic of type-II critical solutions exhibiting discrete
self-similarity, $2M(t,r)/r$ oscillates at higher frequencies and on smaller
spatial scales during the course of an evolution in the critical regime.
As has already been noted, as $l$ increases, the echoing exponent $\Delta_l$
decreases rapidly. This can be observed in Fig.~\ref{raw_data} where the evolution
of the maximum in $r$ is shown as a function of time for four different
values of $L$.

\begin{figure}
\begin{center}
\includegraphics[width=8.0cm,clip=true]{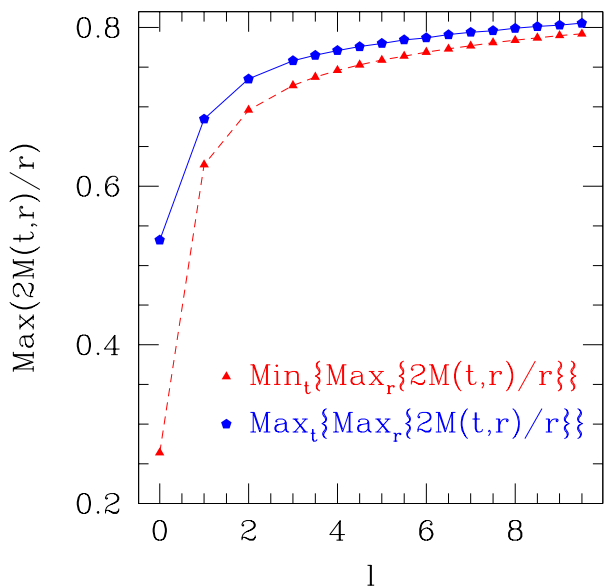}
\end{center}
\caption[Value of $\max_t\left\{\max_r\left\{2M(t,r)/r\right\}\right\}$ in the critical regime as a function of $l$]
{$\max_t\left\{\max_r\left\{2M(t,r)/r\right\}\right\} $ in the critical regime as a function of $l$ (solid line)
and the same for $\min_t\left\{\max_r\left\{2M(t,r)/r\right\}\right\} $ (dashed line).
We see how both the maximum and minimum values of $2M/r$ increase with $l$. On the other hand
the amplitude of oscillation, given by their difference, apparently tends to zero with
increasing $l$.
}
\label{tmr_fig}
\end{figure}

In addition, also in Fig.~\ref{raw_data}, we observe that the maximum and minimum 
values between which the spatial maximum of $2M(t,r)/r$ oscillates increase with $l$
(this fact is shown for all values of $l$ in Fig.~\ref{tmr_fig}) indicating that the critical solutions are becoming
increasingly relativistic as the angular momentum barrier becomes more
pronounced.
The amplitude of the oscillations between these extremal values decreases since
$\min_r\left\{2M(t,r)/r\right\}$
increases more rapidly than $\max_r\left\{ 2M(t,r)/r\right\}$ (see Fig.~\ref{tmr_fig}).

\begin{figure}
\includegraphics[width=8.0cm,clip=true]{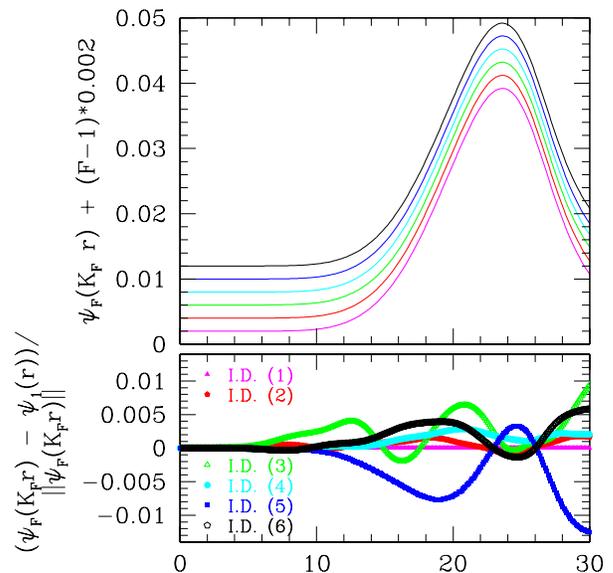}
\caption[Universality Check for l=9.]
{
In the top pane we show the spatial profiles (in the region of self-similarity)
of the scalar field $\psi$ for different families of initial data, but for 
fixed angular momentum parameter $l=9$.
In particular we show the solutions $\psi_F$ 
calculated from initial data types $F={1,...,6}$ (see 
Table \ref{initial_data_table}) at times when $\psi_F$ reaches  maximum amplitude. Each solution is
shifted by an amount proportional to its family number for better visualization, with $F=1$ the bottom curve,
and $F=6$ the top.
The $r$ coordinate is rescaled for each family by a constant
factor $K_F$, which is family dependent, in such a way that the
difference with respect to the profile obtained for the initial data labeled
with $F=1$ (for which we consider $K_1=1$) is minimized. In the bottom pane we 
show the differences
between the rescaled profiles for $F=2,...,6$ and the profile for $F=1$, divided
by the $\ell_2$ norm of the solution.
The maximum relative difference is of the order of a few percent, providing
strong evidence that the critical solution is universal.
}
\label{univ_l=9_fig}
\end{figure}

The assumption that the critical solutions are independent of the initial
family of initial data implies that the spatial profiles at the same moment during
the oscillation for two different families of initial data are the same up to some rescaling
of the radial coordinate. 
In Fig.~\ref{univ_l=9_fig} we show a check of the universality of the spatial 
profile for the solutions computed with $l=9$. Specifically we compare the spatial
profiles at times $T_n$, times at which the local maximum in time 
is achieved during criticality, for different families $F$ of initial data 
$F=1,...,6$ given in Table~\ref{initial_data_table}. In order to compare 
profiles we rescaled the radial coordinate by a constant $K_F$, which depends
on the family of initial data. 
These constants
are chosen in such a way that the $\ell_2$-norm\footnote{The $\ell_2$-norm of a vector u defined
as $||u|| = \sqrt{\sum_{i=1}^N u_i^2/N}$.}
of the difference of the 
profiles with respect to the one with $F=1$, which is considered to have $K_1=1$, 
are minimized. We observed that the maximum of the relative difference, i.e.~the 
difference divided by the $\ell_2$-norm of the solution, is of the order of a few percent,
providing strong evidence for universality.
Similar differences have been observed for other values of the angular momentum parameter.

\begin{figure}
\includegraphics[width=8.0cm,clip=true]{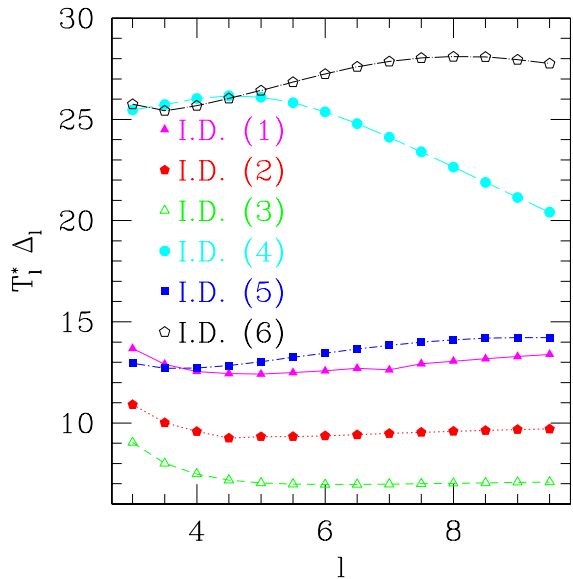}
\caption[$T^\star_l \Delta_l$ as a function of $l$.]
{$T^\star_l \Delta_l$ as a function of $l$. The fact that these products
remain finite as $T^\star_l \to \infty$ and $\Delta_l \to 0$ is
evidence that the critical solutions tend to a {\em periodic} solution
in the limit $l \to \infty$.
\label{periods_fig}
}
\end{figure}

Empirically, we have also found that, as we increase $l$
within a family of initial data, although $\Delta_l \to 0$ and
$T^\star_l \to 0$, the product $T^\star_l \Delta_l$
appears to asymptote to a finite value.
Note that ostensibly this product is family dependent
(see Fig.~\ref{periods_fig}), but again that all DSS type-II critical solutions
are universal only up to a global scale transformation $(r,t) \to (kr,kt)$,
with $k$ an arbitrary positive constant.
Choosing $k=k(l)$ for each of the families so that $\max_r\left\{2 M(t,r)/r\right\}$ 
is attained at some fiducial radius $r_0$, and considering the case $l=10$,
we find that the normalized asymptotic oscillation frequency, $f_0$, defined by
\begin{equation}
\label{f_0}
f_0 = r_0 / (T^\star \Delta) = 4.35 \pm 0.01
\end{equation}
agrees for all families to better than 1\%.
Again, the quoted uncertainty is estimated from the
variation of $f_0$ across the different families of initial data.
We note that for $l=10$ the near-critical solution
stays at a near-constant radial position;
our spatial resolution is insufficient to resolve the small
changes associated with the extremely small value of $\Delta_l$. The radial location of
$\max_r\left\{2 M(t,r)/r\right\}$ in this regime is
the value of $r_0$ that we have used in~(\ref{f_0}).

We also note that the observation that $f_0$ is apparently well defined and
unique (up to the usual rescalings associated with type-II critical solutions),
is consistent with the empirical observation that as $l$ increases, the
critical solution becomes ever closer to a {\em periodic} solution.  In
particular, for a periodic solution we have $\Delta \to 0$, and then
\begin{eqnarray}
T_n-T^\star &=& \left(T_0 - T^\star \right) e^{n \Delta} \approx
(T_0 - T^\star)\left(1+n\Delta \right) \nonumber\\
&\approx& -\left(T^\star \Delta\right)n-T^\star,
\end{eqnarray}
where $T_0$ represents the loosely defined time demarking the onset
of the critical regime (and whose precise value is clearly irrelevant in
the limit $T^\star\to\infty$)
which implies that the maximal value is attained at times $T_n$:
\begin{equation}
T_n = -\left(T^\star \Delta \right) n.
\end{equation}

\begin{figure}[hp]
\includegraphics[width=8.0cm,clip=true]{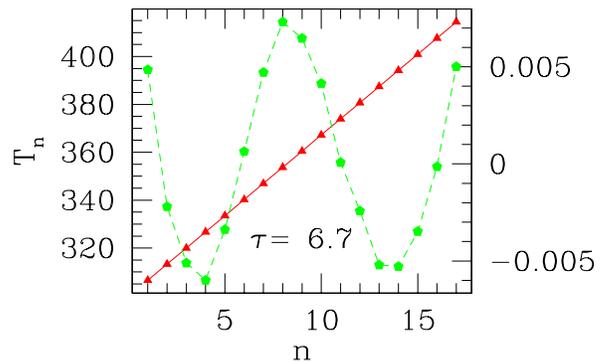}
\caption[Fit to periodicity of the times at which $\max_r\left\{ 2M/r(t,r)\right\}$ attains its maximum.]
{
Fit of the times $T_n$ at which $\max_r\left\{ 2M(t,r)/r\right\}$
reaches its maximum in time (triangles, left scale)
assuming a periodic ansatz.
Initial data type $F=1$  was used with angular momentum parameter $l=10$.
We also plot the residuals of each data point with respect to the best fit
(pentagons, right scale).
}
\label{l=10_linear_fig}
\end{figure}

\begin{figure}
\includegraphics[width=8.0cm,clip=true]{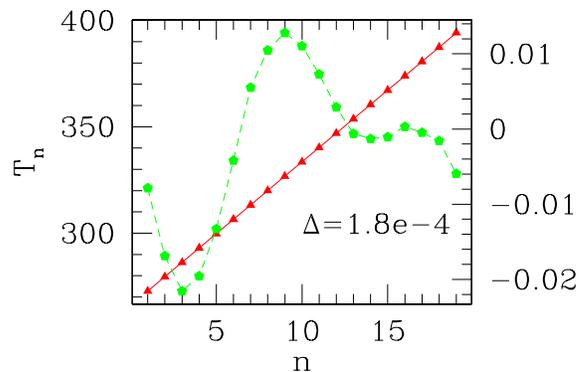}
\caption[Fit to DSS of the times at which $\max_r\left\{2M(t,r)/r\right\}$ attains its maximum.]
{
Fit of the times $T_n$ at which $\max_r\left\{2M(t,r)/r\right\}$
reaches its maximum in time (triangles, left scale)
assuming a self-similar ansatz.
As in the previous plot, initial data type $F=1$ was used with angular momentum parameter $l=10$.
Again, we also plot
the residuals  of each data point with respect to the best fit
(pentagons, right scale).
Notice that the errors in the fit are of the same order
as the errors in the fit that assumes periodicity (Fig.~\ref{l=10_linear_fig}),
indicating that from our numerical results we are unable to
distinguish between the two types of solutions for $l\ge10$.
}
\label{l=10_dss_fig}
\end{figure}

As shown in Figs. \ref{l=10_linear_fig} and \ref{l=10_dss_fig}, from our calculations for $l=10$,
we cannot ascertain whether the solution is discretely
self-similar with $\Delta_l$ very small ($< 0.0002$),
or periodic with period $\tau=T^\star \Delta$.

Naively at least, we expect that for $l > 10$,
distinguishing between discrete self-similarity and periodicity would
become even more difficult.  However, it is worth noting that for $l=20$ we
have {\em not} yet seen evidence for (almost)-periodicity, with period
$T^\star \Delta$,  but have instead seen a more complicated structure
near criticality that is not yet understood.
                                                                                            
\section{Conclusions}
\label{sec:discussion}

In this paper, we have discussed the results for a model that 
incorporates some of the effects of angular momentum in the context
of critical gravitational collapse. A new family of
spherically-symmetric critical solutions,
(black hole threshold solutions)
labelled by an angular momentum parameter, $l$, has been found.
These solutions have similar properties to those for the $l=0$
case originally studied in~\cite{Choptuik:jv}: specifically, the
solutions exhibit  discrete self-similarity, and have scaling laws
for the values of dimensionful quantities in evolutions close to criticality.
We have calculated the $l$-dependence of the echoing exponents
$\Delta_l$, and the mass-scaling exponents
$\gamma_l$, finding that both decrease rapidly with increasing $l$,
(at least up to $l\approx 10$). Moreover,  we have argued that as $l$
increases, the critical solution approaches a periodic evolution.

Together with the results of~\cite{hlcp}, our findings suggest that certain models 
of collapse may generically admit countable infinities of critical 
solutions, each member of which can be characterized by distinct near-origin
regularity conditions (such as (\ref{psi_fall_off}-\ref{pp_fall_off}))  that are preserved by dynamical evolution.
                                                                                            
As we explained in the introduction, we expect that $\gamma_l = 1/\lambda_l$ where
$\lambda_l$ is the Lyapunov exponent associated with the single unstable mode of the 
critical solution for angular momentum parameter $l$. Therefore since 
$\gamma_l \rightarrow 0$ with increasing $l$, we apparently have 
$\lambda_l \rightarrow \infty$. This has the interpretation of increased stability of 
the critical solution for increasing $l$, i.e.~the period of time that a solution
can remain close to criticality (for a fixed amount of fine tuning) increases with $l$.
We believe that this can be interpreted as an effect of the angular momentum barrier 
which (partially) stabilizes the collapse to black hole formation.

\begin{acknowledgments}
It is our pleasure to thank the rest 
of the members of the numerical relativity group at
the University of British Columbia and the members of the Hearne Institute 
in the Department of Physics at LSU for many useful discussions.
In addition, special thanks go to A.~Nagar and L.~Lehner for reading this document.
This research was supported by NSERC, the Canadian Institute for Advanced Research and the 
Government of the Basque Country through a fellowship to I.O.
Most of the calculations were performed on the  
{\tt vn.physics.ubc.ca} Beowulf cluster, which was funded by the 
Canadian Foundation for Innovation, and the British Columbia Knowledge 
Development Fund.
\end{acknowledgments}

\appendix
\section{}
\label{appendixA}
\def\bea{\begin{eqnarray}}
\def\eea{\end{eqnarray}}

\def\ket#1{\left\vert\left. #1 \right\rangle\right.}
\def\bra#1{\left\langle\left. #1 \right\vert\right.}

We wish to show that the stress energy tensor ${\mathcal T}^{(l)a}{}_b=\sum_m {T^{(lm)}}^a{}_{b}$
is independent of $\theta $ and $\phi$, where 
${T^{(lm)}}^a{}_{b}$ is the stress energy tensor associated with the
solution $\psi^{(l)}(t,r)Q_{lm}(\theta,\phi)$, with the same function
$\psi^{(l)}(t,r)$,
for each value of $m$.  For the scalar field, the tensor $T_{a b}$ can be
written in terms of the solutions to the wave equation, $\Psi$, as
\bea
T^{a}{}_{b} = g^{a c}\Psi_{,c}\Psi_{b}- {1\over 2}
\delta^a{}_b g^{dc}\Psi_{,c}\Psi_{d} \, ,
\eea
and if 
\bea
\sum_m
M^a{}_b&=&g^{ac}{(\psi^{(l)}(t,r)Q_{lm}(\theta,\phi))}_{,c}\nonumber\\
&&\times {(\psi^{(l)}(t,r)Q_{lm}(\theta,\phi))}_{,b}
\eea
is independent of $\theta,~ \phi$, then so is ${\mathcal T}^{(l)a}{}_b$. 

Using the definition of the $Q_{lm}$ this can be written in terms of the
$Y_{lm}$ as
\bea
\sum_m
M^a{}_b&=&g^{ac}{(\psi^{(l)}(t,r)Y^*_{lm}(\theta,\phi))}_{,c}\nonumber\\
&&\times{(\psi^{(l)}(t,r)Y_{lm}(\theta,\phi))}_{,b}.
\eea

We can write this in terms of the Green's function
\bea
P(\theta,\phi,\theta',\phi') = \sum_m
Y^*_{lm}(\theta,\phi)Y_{lm}(\theta',\phi')
\eea
in the limit as $\theta'\rightarrow\theta$ and $\phi'\rightarrow\phi$.

In bra-ket notation, this is just the operator
\bea
P=\sum_m \ket{lm}\bra{lm}
\eea
which commutes with all of the angular momentum operators. 
\bea
[ L_z,P ]&=& \sum_m [ L_z, \ket{lm}\bra{lm} ] \nonumber\\
 &=& \sum_m (m
\ket{lm}\bra{lm} -\ket{lm}\bra{lm} m)\nonumber\\ &=&0,
\eea
\bea
[ L_x+iL_y,P ]&=&\sum_m\left( \sqrt{l(l+1)-m^2-m}\ket{lm+1}\bra{lm}\right. \nonumber \\ 
   &&\left. - \ket{lm}\left((L_x-iL_y)\ket{lm}\right)^\dagger\right)\nonumber \\
   &=& \sum_m\left( \sqrt{l(l+1)-m^2-m}\ket{lm+1}\bra{lm}\right. \nonumber\\
   && \left. - \sqrt{l(l+1)-m^2+m}\ket{lm}\bra{lm-1}\right)\nonumber\\ 
&=&0.
\eea
Thus $\sum_m Y^*_{lm}(\theta,\phi)Y_{lm}(\theta',\phi') $ must be a function of
the only rotation invariant function of $\theta,\phi,\theta',\phi'$,
which is the angle $\Theta$ defined by
\bea
\cos(\Theta)=
\cos(\theta)\cos(\theta')+\sin(\theta)\sin(\theta')\cos(\phi-\phi') .
\eea
$\Theta$ is the angle between the two unit vectors with directions $\theta,\phi$ and
$\theta',\phi'$ respectively.
Since $P$ depends only on $\Theta$ we can choose $\theta=0$ to evaluate it,
which gives
\bea
P(\theta,\phi,\theta',\phi') &=&
\sum_m Y^*_{lm}(\theta,\phi)Y_{lm}(\theta',\phi') \nonumber\\ &=&
Y^*_{l0}(0,0)Y_{l 0}(\Theta,0)\nonumber\\ &=& {2l+1\over 4\pi}P_l(\cos(\Theta)).
\eea

The various components of the tensor $M$ are of three types: ones with no
derivatives with respect to $\theta$ or $\phi$ (eg $M_{tt}$), those with
one derivative, (for example $M_{t \theta}$) and those with two (eg, $M_{\theta\theta}$). 
The ones with no derivatives will be functions of
$\lim_{\theta',\phi'\rightarrow \theta,\phi}P= \sqrt{(2l+1)/(4\pi)}\,P_l(1)$
which is clearly independent of $\theta,\phi$. 
The terms with one $\theta,\phi$ derivative will be functions of 
\begin{eqnarray}
\lim_{\Theta\rightarrow 0}\partial_{\theta,\phi} P_l(\cos(\Theta))&=&
\lim_{\Theta\rightarrow 0}
P_l'\{\sin(\Theta),\sin^2(\theta)\sin(\phi-\phi')\}\nonumber\\&=&0,
\end{eqnarray}
and similarly the term 
\bea
M_{\theta\phi} &\propto& \lim_{\Theta\rightarrow 0}\partial_\theta \partial_{\phi'}P_l(\cos(\Theta))\propto \lim_{\phi\rightarrow\phi'}\sin(\phi-\phi')\nonumber\\ &=&0.
\eea
Thus the only two terms remaining are
\bea
M_{\theta\theta}&\propto&
\lim_{\theta\rightarrow\theta'}\partial_\theta\partial_{\theta'}P_l(\cos(\theta-\theta')) \nonumber \\
&=&P_l'(\cos(0))(-\cos(0))(-1)\nonumber\\ &=&P_l'(1)\,,\\
M_{\phi\phi'}&\propto& P_l'(1)\sin(\theta)^2\,.
\eea

Thus the non-zero components   of the tensor $M$ are 
$M_{tt},~M_{tr},~M_{rr},~M_{\phi\phi}=\sin^2(\theta)M_{\theta\theta}$,
with only $M_{\phi\phi}$ having $\theta$ dependence.
Thus,  $M^a{}_b$ will be independent of $\theta,\phi$ and therefore 
so will ${\mathcal T}^{(l)a}{}_b$, as required.

In addition, from the equations of motion for the individual fields $\Psi_{lm}$, each of the energy momentum tensors for
given $l,m$ is conserved in the overall spherically symmetric spacetime.
Thus, so is their sum over $m$ for any given $l$, and we have
\bea
{\mathcal T}^{(l)a}{}_{b;a}=0 \, .
\eea

%

\end{document}